# Potentials of individual atoms by convergent beam electron diffraction

*Tatiana Latychevskaia*[1,2], *Colin Robert Woods*[3,4], *Yi Bo Wang*[3,4], *Matthew Holwill*[3,4], *Eric Prestat*[4], *Sara Mustafi*[1,2], *Sarah J. Haigh*[3,5], *K. S. Novoselov*[3,6,7,8]

[1]*Physics Department, University of Zurich, Switzerland*

[2]*Paul Scherrer Institute, Forschungsstrasse 111, 5232 Villigen, Switzerland, Switzerland*

[3]*National Graphene Institute, University of Manchester, Oxford Road, Manchester, M13 9PL, UK*

[4]*School of Physics and Astronomy, University of Manchester, Oxford Road, Manchester, M13 9PL, UK*

[5]*Department of Materials, University of Manchester, Oxford Road, Manchester, M13 9PL, UK*

[6]*Department of Materials Science and Engineering, National University of Singapore, Singapore, 117575, Singapore*

[7]*Institute for Functional Intelligent Materials, National University of Singapore, 117544 Singapore*

[8]*Chongqing 2D Materials Institute, Liangjiang New Area, Chongqing, 400714, China*

*Corresponding author: tatiana.latychevskaia@psi.ch*

## Abstract

In convergent beam electron diffraction (CBED) on two-dimensional (2D) materials, the intensity distributions within the individual CBED spots map the local atomic arrangements within the probed region. In this study we demonstrate that the average intensities within the CBED spots essentially depend on the scattering parameters by a single atom, thus, offering the possibility of the direct measurement of such parameters. Scattering potential of an individual atom can be approximated by the Gaussian function and its parameters, the standard deviation and the maximal phase shift, can be recovered from the ratio of the intensities in the zero- and higher orders spots in CBED pattern. In order to demonstrate this, we simulated CBED patterns and extracted the atomic scattering parameters from such patterns. In the simulated examples, no weak phase object approximation is applied and the proposed method provides accurate results even for materials with large phase shifts up to 1.5 rad, as for example tungsten. The scattering parameters recovered from the experimental CBED patterns of graphene and twisted graphene-hBN structure show good agreement with the theoretically obtained values.

## 1. Introduction

Quantitative estimation of electron-atom interaction is the key for understanding and interpreting electron images of biological and solid state samples. The whole analysis of biological samples in cryo-electron microscopy is built on using contrast transfer function, which is only valid in the weak phase object approximation (WPOA). The latter holds for light atoms constituting biological samples (H, C, O, N) at typical electron microscope energies (80 - 300 keV). The WPOA also holds for thin two-dimensional (2D) crystals such as graphene or hexagonal boron nitride (hBN). "Weak phase" means that the electron passing by atomic potential will gain an additional phase shift that is much smaller than 1 rad and the transmission function of the sample can be expanded into a Taylor series: $t(x,y)=\exp\left[i\varphi(x,y)\right]\approx 1+i\varphi(x,y)$, leading to linear dependency of the image contrast on the phase shift, which significantly simplifies the image analysis. The phase shift is given by $\varphi(x,y)=\sigma v_z(x,y)$, where $\sigma$ is the interaction parameter and $v_z(x,y)$ is the projected potential of the atom. In simulations, $v_z(x,y)$ is typically modelled using parameters tabulated in the literature [1], but little attention has been paid to recovery of the exact distribution of $v_z(x,y)$ from experimental images. Electron ptychography, a state-of-art resolution-record electron imaging technique, is currently celebrated for achieving a resolution record of 0.4 Å, that is only limited by thermal vibration of atomic lattice [2, 3]. However, quantitative estimation of the phase shifts of the recovered atomic distribution has not been addressed in detail. In fact, the following controversy exists: on one hand, when theory predicts a weak phase shift, the WPOA can be applied, and a simplified data analysis is performed leading to the reconstructed phase shifts that are weak. On the other hand, the reconstructed atomic phase shifts could fall into category of weak phases only because the WPOA and the corresponding simplification were made during the reconstruction procedure. So, the question is, what are the true phase shifts of individual atoms, and, as a consequence, what are the projected potentials of such atoms? Another important questions is, whether both the phase shift and projected potential can be reconstructed directly from a diffraction pattern without a need for applying the WPOA which might bias the output results? In this study we use a 2D graphene as a test sample to investigate the electron-matter interaction. We quantitatively evaluate the phase shifts and projected potential of individual atoms from the simulated and experimental CBED pattern and compare the obtained values to the theoretical values.

## 2. Methods or Experimental and Computational Details

### 2.1 CBED arrangement

Convergent beam electron diffraction (CBED) has routinely been applied for three-dimensional (3D) crystals since its invention in 1939 [4]. Recently, CBED has been demonstrated for 2D crystals, where the data analysis is straight forward and the information about lattice defects [5, 6], and interlayer distances in bilayer structures can be directly retrieved from interference patterns in individual CBED spots [6, 7]. A diagram for a CBED experiment is shown in Fig. 1a. In conventional electron diffraction a parallel electron beam is probing the sample and the resulting diffraction pattern exhibits sharp Bragg peaks which occupy a few pixels on the detector. The proposed here analysis requires recording of the exact intensity values of all diffraction peaks, including the zero-order peak, which is nearly impossible to realize in a conventional diffraction mode. Firstly, because the intensity values of individual sharp diffraction peaks cannot be accurately measured due to the finite pixel size, so that even the diffraction peaks of the same order would exhibit different intensity. (2) Secondly, the zero-order peak intensity is rarely accessible in an electron diffraction experiment because it is blocked due its high intensity. In CBED, a convergent electron beam is probing the sample which turns the sharp diffraction peaks into finite size spots. The other advantage of CBED of 2D materials over conventional diffraction is that the intensity distribution in a CBED spot can be directly mapped to the atomic distribution in the probed region. When CBED disks are not overlapping, the intensity distribution within a CBED disk can be analysed as a hologram and the 3D atomic positions can be retrieved. A CBED with non-overlapping disks is acquired when the convergence semiangle $\alpha$ is selected to be smaller than half of the Bragg diffraction angle (19.6 mrad for graphene). In addition, for quantitative analysis, presented in this paper, the detector settings should be selected in such a way, that all CBED disks are simultaneously recorded in one CBED pattern, that is, including the zero-order CBED spot.

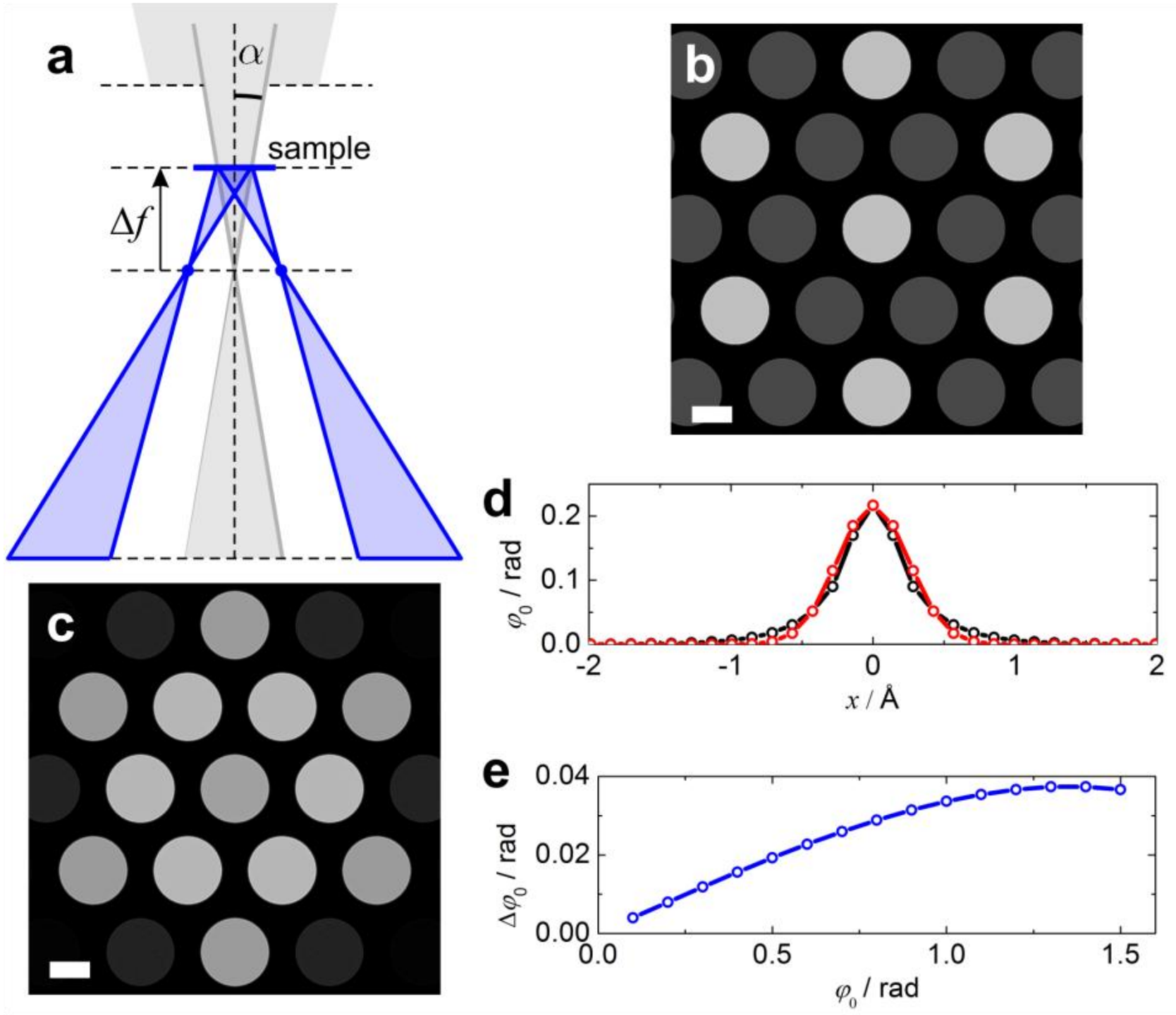

Fig. 1 Convergent beam electron diffraction of a 2D crystal. (a) Scheme of experimental arrangement, $\alpha$ is the convergence semiangle, $\Delta f$ the defocus. (b) CBED pattern of a hexagonal crystal simulated by electron scattered as spherical waves at atomic positions. (c) CBED pattern of graphene simulated by electrons scattered by the projected potential of the atoms. The zero-order spot intensity is reduced by factor 1E-4. (b) and (c) are simulated at 80 keV. Scalebars in (b) and (c) are 2 1/nm. (d) The phase shift caused by a single carbon atom in 80keV electron wave, calculated using Kirkland's parameters (black), with a Gaussian function fit (red). (e) Mismatch between the original and the reconstructed amplitude of the phase shift $\Delta\varphi_0 = \varphi_0^{(\mathrm{rec})} - \varphi_0$.

For a perfectly flat uniform lattice, intensities within the CBED spots are constant, which we will refer to as an "average intensity in a CBED spot":

$$\left\langle I^{(n)} \right\rangle = \frac{1}{A^{(n)}} \sum_{k_x, k_y} I^{(n)}\left(k_x, k_y\right), \tag{1}$$

where $I^{(n)}\left(k_x, k_y\right)$ is the intensity value at $\left(k_x, k_y\right)$ pixel in ($n$)-th CBED spot, and $A^{(n)}$ is the total area of the CBED spots of ($n$)-th order in pixels.

## 2.2 Projected potential from intensities ratios in a CBED pattern

### 2.2.1 Atoms as point scatterers

In the simplest model, when an electron wave illuminates the sample, each atom acts as a point scatterer and emits a spherical wave; the corresponding simulated CBED pattern is shown in Fig. 1b. The graphene lattice can be represented as sum of two trigonal lattices with a relative shift of $\Delta x = a_0 = 142$ pm, where $a_0$ is the closest atom-to-atom distance in graphene. The wavefront in the far field is given by:

$$U\left(k_x, k_y\right) = U_0\left(k_x, k_y\right)\left[1 + \exp\left(-ik_x \Delta x\right)\right] \tag{2}$$

where the Fourier transform is defined as:

$$U_0\left(k_x, k_y\right) = \iint l_0\left(x, y\right) \exp\left[-i\left(k_x x + k_y y\right)\right] \mathrm{d}x\mathrm{d}y, \tag{3}$$

and $l_0\left(x, y\right)$ is the exit wave behind one trigonal lattice. The intensity is given by

$$I\left(k_x, k_y\right) = 2I_0\left(k_x, k_y\right)\left[1 + \cos\left(k_x \Delta x\right)\right], \tag{4}$$

where $I_0\left(k_x, k_y\right)$ is the distribution of intensity for one trigonal lattice monolayer. From Eq. (4), the intensities of the individual CBED spots of different orders can be evaluated as intensity values at the $\left(k_x, k_y\right)$ values corresponding to the center of a CBED spot: $\left\langle I^{(0)} \right\rangle = 4I_0$, $\left\langle I^{(1)} \right\rangle = I_0$, $\left\langle I^{(2)} \right\rangle = 4I_0$. The intensity of CBED spots of different orders relate as $I^{(0)} : I^{(1)} : I^{(2)} \rightarrow 4:1:4$, this relation is also observed in the simulated CBED pattern shown in Fig. 1b.

### 2.2.2 Electron scattering on atomic potential

A more realistic model of electron scattering on a sample utilizes scattering of electrons by the potentials of individual atoms in 2D crystal. In this model, the total transmission function of the sample is written as

$$t\left(x, y\right) = \exp\left[i\Phi\left(x, y\right)\right] = \exp\left[i\varphi\left(x, y\right) \otimes l\left(x, y\right)\right] \tag{5}$$

where $\varphi\left(x, y\right)$ is a phase shift by a single atom in the lattice, and $l\left(x, y\right)$ is the function describing the positions of individual atoms in the lattice. The sample phase shift is given by $\Phi\left(x, y\right) = \sigma V_z\left(x, y\right)$, where $V_z\left(x, y\right)$ is the sample projected potential. The interaction parameter $\sigma$ is given by $\sigma = \frac{2\pi m e \lambda}{h^2} = \frac{2\pi e}{h\upsilon}$, where $m$ is the relativistic mass of the electron, $e$ is an elementary charge, $\lambda$ is wavelength of the electrons, $\upsilon$ is the electron velocity, and $h$ is the Planck's constant. $V_z\left(x, y\right) = v_z\left(x, y\right) \otimes l\left(x, y\right)$, where $v_z\left(x, y\right)$ is the projected potential of a single atom. Projected potential of an individual atom $v_z(x, y)$ can be modelled using parameters tabulated in the literature [1, 8]. Using these, the maximal phase shift of an electron beam of 80 keV calculated at radial position of 0.1 Å amounts to: 0.217 rad for carbon and 0.214 rad for hBN (calculated as an average of 0.190 rad for B and 0.238 rad for N).

### 2.2.3 Weak phase object approximation (WPOA)

A CBED pattern of graphene simulated via scattering of electrons by the atomic potentials is shown in Figure 1c. It was calculated using the transmission function given by Eq. (5), without approximations. One significant difference between the CBED patterns shown in Fig. 1 b and c is that the zero-order CBED spot is about $10^4$ times more intense when electron scattering on the atomic potentials is taken into account. Mathematically, it can be explained as follows. For weak phase, the transmission function given by Eq. (5) can be expanded in Taylor series as

$$t(x, y) \approx 1 + i\Phi(x, y) \tag{6}$$

where term "1" is much larger than the second term. $t(x, y) = 1$ corresponds to the situation without any sample and thus describes the strong signal in the zero-order CBED spot, and only in the zero-order CBED spot. The second term in Eq. (6) gives rise to a periodical signal in Fourier space, that is CBED spots of all orders, including the zero-order CBED spot; however, the signal in the zero-order CBED spot, which comes from the second term is negligibly small when comparing to the contribution from the unscattered wave described by the first term. In all simulations shown in this study, no WPOA is applied, and the transmission function is calculated by Eq. (5).

### 2.2.4 Gaussian distribution phase shift parameters evaluated from CBED pattern

The intensity within the zero-order CBED spot is given mainly by the unscattered probing wave, the contribution from the scattered wave is about $10^3$ times smaller. The signals within higher-order CBED spots are given only by the scattered wave, and thus solely determined by the scattering potential of individual atoms in the lattice. Thus, neglecting the contribution from the scattered wave, the intensity of the probing beam can be determined from the averaged intensity in the zero-order CBED spot. Next, knowing the exact quantitative distribution of the probing wave, in particular its amplitude, the parameters of the scattering potential of individual atoms can be recovered from the intensities in the first- and higher order CBED spots. In the next subsections we will show that the phase shift introduced by an individual atom can be modelled as a Gaussian function and we derive the equations and demonstrate how the amplitude and standard deviation of this Gaussian function can be retrieved from the intensities of zero- and higher-orders CBED spots.

## 2.3 Gaussian distribution phase shift parameters evaluated from the CBED pattern

### 2.3.1 Phase shift modelled as a Gaussian distribution

Phase shift, which an electron wave acquires by interaction with an individual atom, can be accurately modelled using the parameters tabulated in refs [1, 8]. In this study, the phase shift is modelled in the form of a Gaussian function:

$$\varphi(x,y) = \varphi_0 \exp\left(-\frac{x^2+y^2}{2\sigma_\mathrm{p}^2}\right), \tag{7}$$

with amplitude $\varphi_0$ and standard deviation $\sigma_\mathrm{p}$. This modelling allows us to freely change the parameters and see whether the same values of the parameters are reconstructed by the proposed method. Fig. 1d shows the phase shift of a single carbon atom calculated using Kirkland's parameters, and a superimposed fit by a Gaussian function such as Eq. (7) with $\varphi_0 = 0.217$ rad and $\sigma_\mathrm{p} = 0.252$ Å.

For larger $\varphi_0$, and the coordinates $(x,y)$ in the vicinity of the atom, the following Taylor series expansion can be applied giving: $t(x,y) = \exp\left[i\varphi(x,y)\right] = \left\{\exp\left[i\varphi(x,y)/\tau\right]\right\}^\tau \approx \left[1+i\varphi(x,y)/\tau\right]^\tau \approx 1+i\varphi(x,y)$, where $\tau \gg 1$ is some constant. The simulations showed that Fourier transform of the distribution $\exp\left[i\varphi_0 \exp\left(-\frac{x^2+y^2}{2\sigma_\mathrm{p}^2}\right)\right]$ indeed gives a sharp peak in the center of the spectrum due to the constant terms and it gives a Gaussian-distributed signal due to the Fourier transform of $i\varphi_0 \exp\left(-\frac{x^2+y^2}{2\sigma_\mathrm{p}^2}\right)$, see Appendix A.

The Fourier transform of a Gaussian function is also a Gaussian function, and the Fourier transform of $\varphi(x,y)$ is given by:

$$\mathrm{FT}(\varphi) = \iint \varphi(x,y) \exp\left[-i\left(xk_x + yk_y\right)\right] \mathrm{d}x\mathrm{d}y = 2\pi\sigma_\mathrm{p}^2 \varphi_0 \exp\left(-\frac{k_x^2+k_y^2}{2\sigma_\mathrm{p,F}^2}\right) \tag{8}$$

with

$$\sigma_\mathrm{p,F} = \frac{1}{\sigma_\mathrm{p}}. \tag{9}$$

### 2.3.2 Finding the standard deviation

Using the Taylor series expansion (in the WPOA or as described above for strong phase shifts), the intensity distribution of a CBED pattern can be written as;

$$I\left(k_x, k_y\right) \approx \left|\mathrm{FT}\left[p\left(1 + i\varphi \otimes l\right)\right]\right|^2, \tag{10}$$

where $p$ is the complex-valued distribution of the probing wave, and FT means Fourier transform.

The average intensity of the zero-order CBED spot is mainly given by the unscattered wave (term "1" in Eq. (10)):

$$I^{(0)} \approx \left|\mathrm{FT}\left(p\right)\right|^2. \tag{11}$$

The signal in the higher-order CBED spots is provided by the scattered wave only. For a Gaussian-distributed phase shift of individual atoms, the average intensity in a higher-order CBED spot can be found via analytical solution:

$$\left\langle I^{(n)} \right\rangle \approx \left|\mathrm{FT}\left(p\varphi\right)\right|^2 = p_0^2 \left(2\pi\sigma_\mathrm{p}^2\right)^2 \varphi_0^2 \exp\left(-\frac{k_x^2 + k_y^2}{\sigma_\mathrm{p,F}^2}\right) = p_0^2 \left(2\pi\sigma_\mathrm{p}^2\right)^2 \varphi_0^2 \exp\left[-\frac{\left(k^{(n)}\right)^2}{\sigma_\mathrm{p,F}^2}\right], \tag{12}$$

where $k^{(n)}$ is the coordinate of the center of the *n*-th order CBED spot and $p_0$ is the amplitude of the probing wave. From Eq. (12) it can be seen that the standard deviation $\sigma_\mathrm{p,F}$ can be directly recovered by comparing the averaged intensities in the first- and the second order CBED spots:

$$\frac{4\left\langle I^{(1)} \right\rangle}{\left\langle I^{(2)} \right\rangle} = \exp\left[-\frac{\left(k^{(1)}\right)^2 - \left(k^{(2)}\right)^2}{\sigma_\mathrm{p,F}^2}\right] = \exp\left\{-\left[\left(k^{(1)}\right)^2 - \left(k^{(2)}\right)^2\right]\sigma_\mathrm{p}^2\right\}, \tag{13}$$

where the factor 4 accounts for the difference in the intensity due to the hexagonal lattice, without projected potential, as discussed above.

### 2.3.3 Finding the amplitude

The Fourier transform of the phase shift function $\varphi\left(x, y\right)$ in Cartesian coordinates is given by:

$$\mathrm{FT}\left(\varphi\right) = -\frac{i}{\lambda z} \iint \varphi\left(x, y\right) \exp\left[-\frac{2\pi i}{\lambda z}\left(xX + yY\right)\right] \mathrm{d}x\mathrm{d}y = -\frac{2\pi i}{\lambda z} \sigma_\mathrm{p}^2 \varphi_0 \exp\left(-\frac{X^2 + Y^2}{2\sigma_\mathrm{d}^2}\right), \tag{14}$$

where

$$\sigma_\mathrm{d} = \frac{\lambda z}{2\pi\sigma_\mathrm{p}}. \tag{15}$$

The averaged intensity in the zero-th order CBED spot is given approximately by the probing wave alone and amounts to:

$$\left\langle I^{(0)} \right\rangle = I_0 \left( \frac{1}{M} \right)^2 = I_0 \left( \frac{\Delta f}{z} \right)^2 \qquad (16)$$

where $M = z / \Delta f$ is the magnification factor, $\Delta f$ is the defocus distance, $z$ is the sample-to-detector distance (camera length), and $I_0$ is the intensity of the probing wave.

The averaged intensity in a first-order CBED spot of a monolayer depends on the number of the scattering atoms (that is, the number of atoms in the probed region) and on the phase shift distribution parameters as follows:

$$\left\langle I^{(1)} \right\rangle = I_0 N_0 \left[ \frac{2\pi}{\lambda z} \sigma_{\mathrm{p}}^2 \varphi_0 \exp\left( -\frac{r_{\mathrm{d}}^2}{2\sigma_{\mathrm{d}}^2} \right) \right]^2 = 2 I_0 \frac{\pi}{h_0} (\Delta f \tan\alpha)^2 \left[ \frac{2\pi}{\lambda z} \sigma_{\mathrm{p}}^2 \varphi_0 \exp\left( -\frac{r_{\mathrm{d}}^2}{2\sigma_{\mathrm{d}}^2} \right) \right]^2 \qquad (17)$$

where $\sigma_{\mathrm{d}}$ is given by Eq. (15), $r_{\mathrm{d}} = \sqrt{X^2 + Y^2}$ is the coordinate of the center of the CBED spot, $N_0$ is the number of atoms in the probed area, $N_0 = 2 \frac{\pi}{h_0} (\Delta f \tan\alpha)^2$, $h_0$ is the area of one 2D unit cell in the graphene lattice, and the factor 2 accounts for two carbon atoms per one unit cell. Note that $\left\langle I^{(0)} \right\rangle$ and $\left\langle I^{(1)} \right\rangle$ both contain factor $I_0 \left( \frac{\Delta f}{z} \right)^2$, which is eliminated when calculating the ratio $\left\langle I^{(1)} \right\rangle / \left\langle I^{(0)} \right\rangle$:

$$\frac{\left\langle I^{(1)} \right\rangle}{\left\langle I^{(0)} \right\rangle} = 2\gamma^2 \frac{\pi}{h_0} (\tan\alpha)^2 \left[ \frac{2\pi}{\lambda} \sigma_{\mathrm{p}}^2 \varphi_0 \exp\left( -\frac{r_{\mathrm{d}}^2}{2\sigma_{\mathrm{d}}^2} \right) \right]^2, \qquad (18)$$

where $\gamma$ is the number of layers. The ratio $\left\langle I^{(1)} \right\rangle / \left\langle I^{(0)} \right\rangle$ does not depend on the intensity of the probing beam $I_0$, defocus distance $\Delta f$ or the sample-to-detector distance $z$. By using Eqs. (9) and (13), $\varphi_0$ can be determined from the known parameters $\alpha$, $\lambda$, $h_0$, ratio of measured intensities $\left\langle I^{(1)} \right\rangle / \left\langle I^{(0)} \right\rangle$, and $\sigma_{\mathrm{p}}$ determined from the ratio $\left\langle I^{(1)} \right\rangle / \left\langle I^{(2)} \right\rangle$. When the parameters $\sigma_{\mathrm{p}}$ and $\varphi_0$ are known, the number of layers $\gamma$ can be evaluated by comparing the averaged intensities in the zero- and first order CBED spots using Eq. (18).

# 3. Results and Discussion

## 3.1 Parameters reconstructed from simulated CBED patterns

A series of CBED patterns for atomic lattices with different parameters for the phase shift of individual atoms were simulated (as described in Appendix B) and the parameters were recovered by using the equations above. CBED patterns were simulated at defocus values $\Delta f$ = 0.5, 1.0, and 2.0 um, and it was also verified that the recovered standard deviation $\sigma_{\mathrm{p}}$ and amplitude $\varphi_0$ did not depend on the defocus value $\Delta f$ . CBED patterns were calculated for phase shifts with the standard deviation $\sigma_{\mathrm{p}} =$ 0.2 Å and amplitudes $\varphi_0 =$ 0.1, 0.2, ...1.5 radian.

The scattering parameters were evaluated from the simulated CBED patterns. $\sigma_{\mathrm{p}}$ was evaluated from the ratio $\langle I^{(1)} \rangle / \langle I^{(2)} \rangle$ using Eq. (13), and $\varphi_0$ from the ratio $\langle I^{(1)} \rangle / \langle I^{(0)} \rangle$ using Eq. (18). The standard deviation $\sigma_{\mathrm{p}}$ was correctly reconstructed as 0.2 Å from all the CBED patterns. The reconstructed phase shifting amplitudes $\varphi_0$ slightly deviated from the original values, with the mismatch increasing towards higher $\varphi_0$, reaching a mismatch of 0.04 rad for $\varphi_0 = 1.5$ rad, as illustrated in Fig. 1e. This result is quite interesting. $\varphi_0 = 1.5$ rad is a large phase shift, for which the conventional WPOA cannot be applied. The corresponding CBED pattern was simulated without any approximations and the recovered (relatively large) phase shift $\varphi_0$ is almost correct. This confirms that the approximation for strong phase shifts which we derived above can be applied and leads to correct results. The proposed analysis can be applied for determination of scattering potentials of arbitrary atoms which cause even larger phase shifts than required by the conventional WPOA. This could include higher atomic number atoms, such as molybdenum ($\varphi_0 = 0.969$ rad at 80 keV) or tungsten ($\varphi_0 = 1.415$ rad at 80keV), or for other atoms in transition metal dichalcogenide (TMD) structures.

The CBED pattern of a bilayer 2D material structure, where atoms in each layer exhibit $\sigma_{\mathrm{p}} =$ 0.2 Å and $\varphi_0 =$ 0.1 rad, with a twist angle of 2° was simulated. In the analysis, only the areas where the CBED spots from individual layers overlap were considered (regions with overlapping and non-overlapping CBED spots are illustrated in Fig. 2b). For a bilayer system, the average intensity in a CBED spot amounts to the sum of the intensities from individual layers. $\sigma_{\mathrm{p}}$ was determined from the ratio $\langle I^{(1)} \rangle / \langle I^{(2)} \rangle$ using Eqs. (9) and (13); here the intensity in both the first- and second-order CBED spots was higher by factor of 2 due to the presence of two layers. $\varphi_0$ was determined from the ratio $\langle I^{(1)} \rangle / \langle I^{(0)} \rangle$ using Eq. (18); here the two layers led to the intensity in the first-order CBED spots increasing by factor of 2, so that the measured intensity $\langle I^{(1)} \rangle$ needed to be reduced by factor 2 in Eq. (18). The recovered values are $\sigma_{\mathrm{p}} = 0.212$ Å and $\varphi_0 = 0.096$ rad. When comparing to the original values, the errors are small, and can be explained by the intensity values being averaged over the interference patterns in the first- and second-order CBED spots, where the amount of maxima and minima might be not exactly the same so as to get the correct average.

## 3.2 Parameters reconstructed from experimental CBED patterns

### 3.2.1 Monolayer graphene

The experimental CBED pattern of monolayer graphene is shown in Fig. 2a. The samples were prepared and imaged in CBED mode as described in Appendix C, and the obtained CBED patterns were analysed as described in Appendix D. The following parameters of the Gaussian distribution were reconstructed by the proposed method: $\sigma_{\mathrm{p}} = 0.308$ Å and $\varphi_0 = 0.156$ rad, which compares well to the theoretically calculated values mentioned above: $\sigma_{\mathrm{p}} = 0.252$ Å and $\varphi_0 = 0.217$ rad. The main source of the mismatch is the noise, and the presence of the ripples or adsorbates on the graphene surface which lead to the intensity variations within CBED spots, as it can be seen in the CBED pattern shown in Fig. 2a.

### 3.2.2. Bi-layer graphene-hBN

The experimental CBED pattern of twisted bilayer graphene-hBN is shown in Fig. 2b. In this CBED pattern, the contribution from individual layers can be extracted from the regions of non-overlapping disks. The signal in the CBED disks can be easily assigned to the individual layers. The CBED disks observed at higher scattering angles are from the graphene layer (smaller interatomic distance), and the CBED disks observed at lower scattering angles are from the hBN (larger interatomic distance). By using the quantitative analysis, we obtain $\sigma_{\mathrm{p}} = 0.229$ Å and $\varphi_0 = 0.189$ rad for the graphene layer and $\sigma_{\mathrm{p}} = 0.244$ Å and $\varphi_0 = 0.162$ rad for the hBN layer. As predicted by theory, the maximal phase shift is higher for graphene. Quantitatively, the maximum phase shifts for both layers are somewhat smaller than expected from the theory, $\varphi_0 = 0.217$ rad for graphene and $\varphi_0 = 0.214$ rad for hBN.

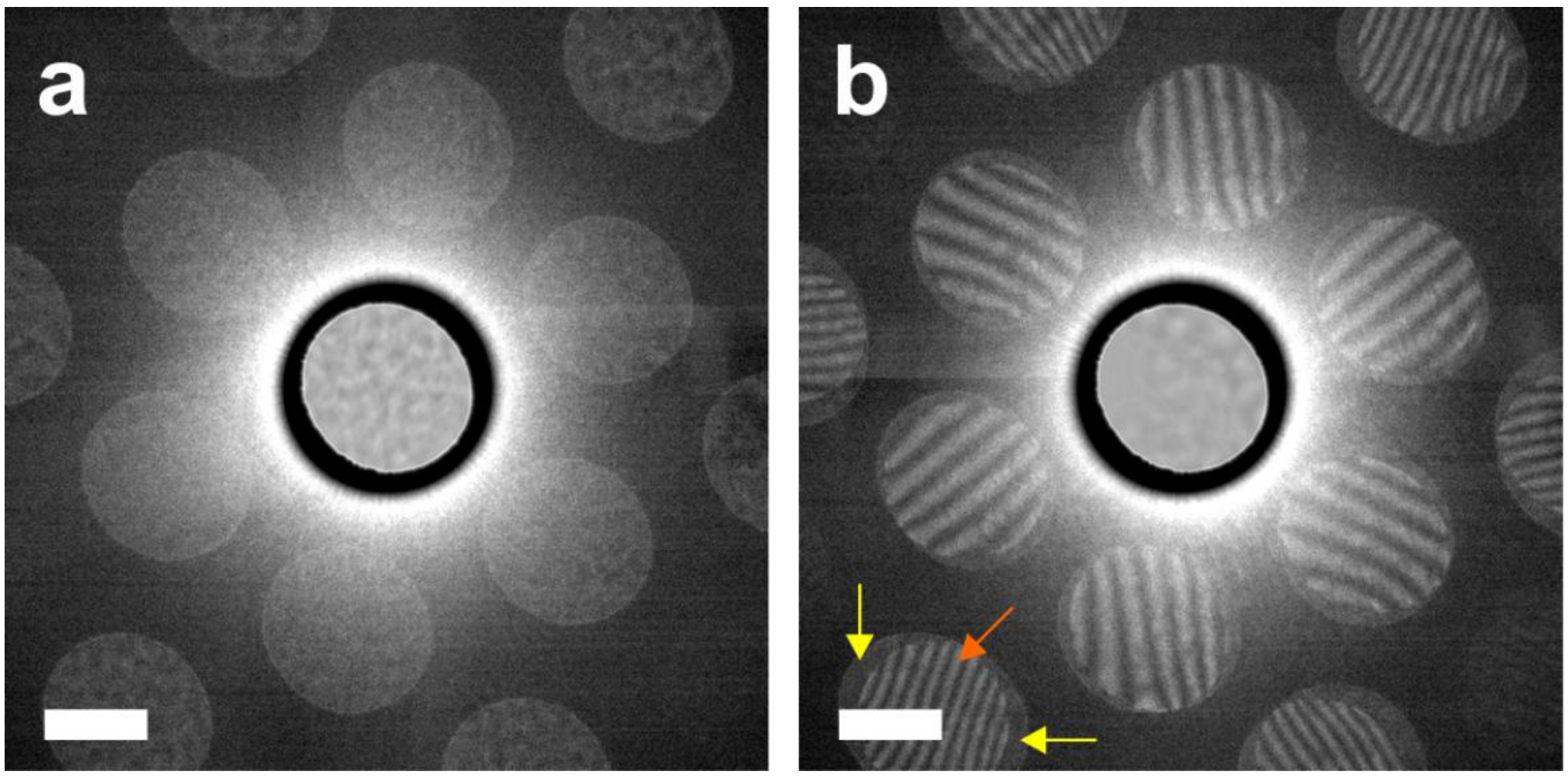

Fig. 2 Experimental convergent beam electron diffraction (CBED) patterns of (a) monolayer graphene and (b) monolayer graphene on monolayer hBN. Both CBED patterns acquired at 80 keV and $\Delta f$ = 1.8 um. The zero-order spot intensity is reduced by factor $10^{-3}$. In (b), a region with overlapping spots is indicated by the orange arrow, regions with non-overlapping spots are indicated by the yellow arrows. Scalebars are 2 1/nm.

For a twisted bilayer system, interference pattern is observed in the regions where the CBED disks overlap [6]. The averaged intensity of such an interference pattern is given by the sum of the averaged intensities from the two layers. By using quantitative analysis on the areas where the CBED disks overlap, the following parameters were recovered $\sigma_{\mathrm{p}} = 0.295$ Å and $\varphi_0 = 0.153$ rad. The reconstructed standard deviation $\sigma_{\mathrm{p}}$ appears larger than determined from the CBED intensity of individual layers and the maximal phase shift $\varphi_0$ is smaller than that recovered for individual layers. This mismatch is similar to the mismatch observed for the simulated CBED pattern of the bilayer system. Similar to the case of the simulated CBED pattern for the bilayer system, the errors here can be explained by errors introduced by averaging over the interference pattern.

## 4. Conclusions

We derived equations for recovery of single atom scattering parameters, the maximum phase shift and the extent of the scattering potential, directly from corresponding CBED pattern and ratio of the intensities in the zero- and higher orders CBED spots. The scattering parameters recovered from simulated and experimental CBED patterns exhibit good agreement with the theoretical values. The simulations confirm that the proposed analysis can also be applied for atoms which scatter strongly and introduce relatively large phase shifts into the probing electron beam. Thus, the proposed data analysis could be utilized for determination of scattering potentials of arbitrary atoms, including atoms in TMD structures.

The main sources of errors are the noise and presence of ripples or adsorbates on the graphene surface which lead to intensity variations within CBED spots. For bilayer twisted systems, the defocus values should be selected to be relatively large so that the interference patterns in the individual CBED spots exhibit several periods and the averaged intensity can be extracted accurately.

Extraction of the number of layers from the electron diffraction data proved to be a challenging task in electron diffraction imaging of van der Waals structures [9, 10]. When the parameters of individual atom phase shift are known, the number of layers can be determined from the ratio of the average intensities in the zero- and first orders.

# 5. Appendices

## A. Fourier transform of function with Gaussian-distributed phase

The following approximation $\exp\left[i\varphi(x,y)\right] = \exp\left[i\varphi_0 \exp\left(-\frac{x^2+y^2}{2\sigma_{\mathrm{p}}^2}\right)\right] \approx 1 + i\varphi_0 \exp\left(-\frac{x^2+y^2}{2\sigma_{\mathrm{p}}^2}\right)$ can be applied even for large $\varphi_0$, as demonstrated from the Fourier transform distribution $u(v,w) = \mathrm{FT}\left\{\exp\left[i\varphi(x,y)\right]\right\}$, shown in Fig. A. 1. The real part of $u(v,w)$ is mainly given by strong peak in the center $\delta(v,w)$ due to $\exp\left[i\varphi(x,y)\right] \approx 1$. Besides the strong peak associated with $\delta(v,w)$ function, the real- and imaginary parts of $u(v,w)$ exhibit signal given by Gaussian distributions, as provided by Fourier transform of $i\varphi_0 \exp\left(-\frac{x^2+y^2}{2\sigma_{\mathrm{p}}^2}\right)$.

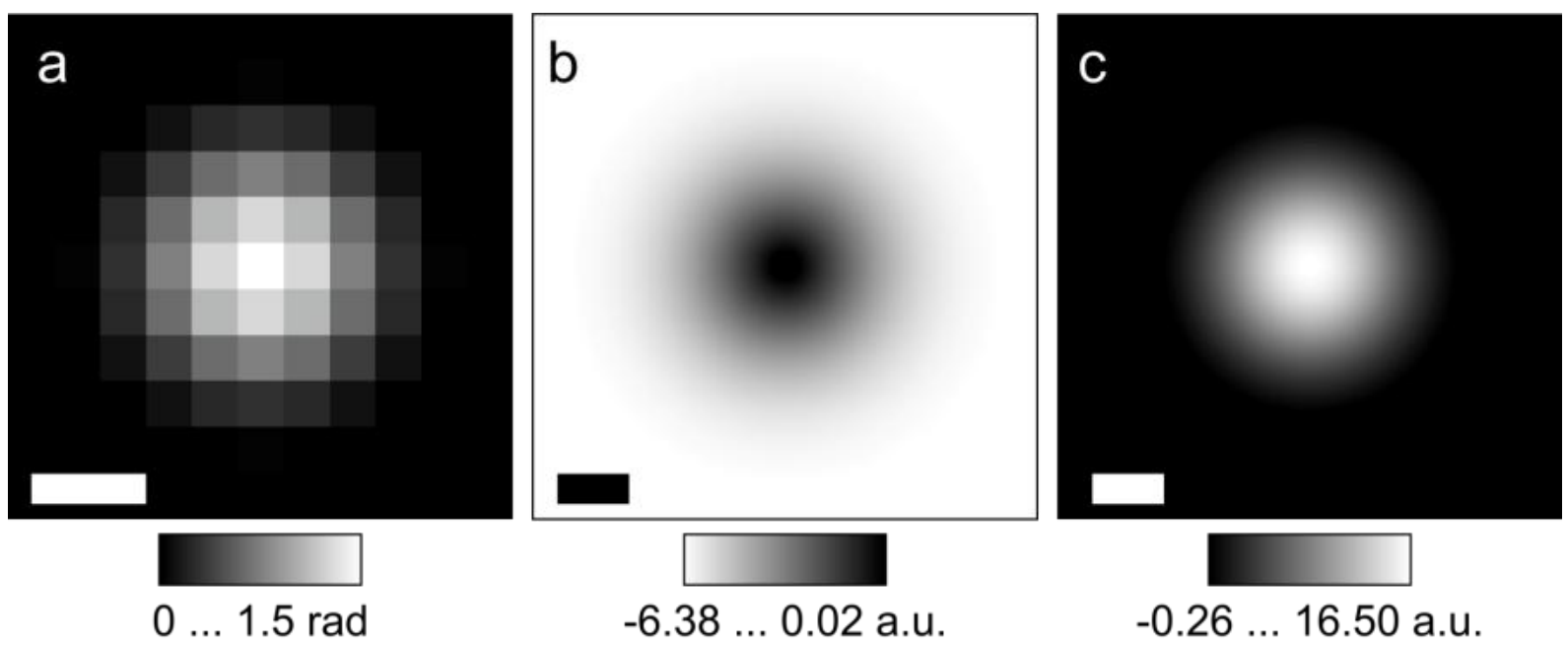

Fig. A. 1. Fourier transform of transmission function with Gaussian-distributed phase. (a) Phase distribution $\varphi_0 \exp\left(-\frac{x^2+y^2}{2\sigma_{\mathrm{p}}^2}\right)$ with $\varphi_0 =$ 1.5 rad, $\sigma_{\mathrm{p}} =$0.2 Å, pixel size = 0.142

Å, central region of 11 × 11 pixel shown. Scalebar is 0.5 Å. (b) Real part of Fourier transform of $\exp\left[i\varphi_0 \exp\left(-\frac{x^2+y^2}{2\sigma_{\mathrm{p}}^2}\right)\right]-1$ , where 1 is subtracted to remove the sharp peak in the center. (b) Imaginary part of Fourier transform of $\exp\left[i\varphi_0 \exp\left(-\frac{x^2+y^2}{2\sigma_{\mathrm{p}}^2}\right)\right]$. Scalebars in (b) and (c) are 10 1/nm.

## B. Simulation of CBED patterns

CBED patterns were simulated as described in refs [6, 11].

**Probing wave distribution.** The probing wave distribution $\psi_0(\vec{r})$ was modelled by simulating the diffraction of a spherical wavefront on a limiting virtual aperture (second condenser aperture) positioned at a plane $\vec{r}_0$:

$$\psi_0(\vec{r}) = \iint a(\vec{r}_0) \frac{\exp(-ikr_0)}{r_0} \frac{\exp(ik|\vec{r}_0 - \vec{r}|)}{|\vec{r}_0 - \vec{r}|} \mathrm{d}\vec{r}_0, \tag{A.1}$$

where $a(\vec{r}_0)$ is the aperture binary function. In the simulations, the integral transform in Eq. (A.1) was calculated directly, without applying a fast Fourier transform (FFT), as

$$\psi_0(p,q) = \sum_{m,n=-N/2}^{N/2-1} a(m,n) \frac{\exp\left(-ik\sqrt{(m\Delta_a)^2 + (n\Delta_a)^2}\right)}{\sqrt{(m\Delta_a)^2 + (n\Delta_a)^2}} \frac{\exp\left(ik\sqrt{(m\Delta_a - p\Delta)^2 + (n\Delta_a - q\Delta)^2}\right)}{\sqrt{(m\Delta_a - p\Delta)^2 + (n\Delta_a - q\Delta)^2}},$$

where *m*, *n*, *p*, *q* are the pixel indices in the aperture (*m*, *n*) and the sample (*p*, *q*) planes, and $\Delta_a$ and $\Delta$ are the pixel sizes in the aperture and the sample planes, respectively. The convergence semiangle of $\alpha$ = 7 mrad was achieved by selecting a virtual exit aperture of diameter 640 nm at a distance of 45 μm from the virtual source plane.

**CBED patterns of monolayer samples** The input data consisted of an array of the coordinates of all the atoms $(x_n, y_n)$. The transmission function of a monolayer was calculated as:

$$t(x,y) = \exp\left[i\sigma v_z(x,y) \otimes l(x,y)\right], \tag{A.2}$$

where $v_z(x,y)$ is the projected potential of an individual atom, $l(x,y)$ is the function describing positions of the atoms in the lattice, and $\otimes$ denotes convolution. The projected potential of a single carbon atom was simulated in the form ref. [1]:

$$v_z(r) = 4\pi^2 a_0 e \sum_{i=1}^{3} a_i K_0\left(2\pi r\sqrt{b_i}\right) + 2\pi a_0 e \sum_{n=1}^{3} \frac{c_i}{d_i} \exp\left(-\pi^2 r^2 / d_i\right),$$

where $r=\sqrt{x^2+y^2}$, $a_0$ is the Bohr radius, $e$ is the elementary charge, $K_0(\ldots)$ is the modified Bessel function, and $a_i, b_i, c_i, d_i$ are parameters that depend on the chemical origin of the atoms and are tabulated elsewhere [1]. In $v_z(r)$, the singularity at $r=0$ is replaced by the value of $v_z(r)$ at $r$ = 0.1 Å. The convolution $v_z(x,y)\otimes l(x,y)$ in Eq. (A.2) is calculated as $\mathrm{FT}^{-1}\left\{\mathrm{FT}\left[v_z(x,y)\right]\mathrm{FT}\left[l(x,y)\right]\right\}$. $\mathrm{FT}\left[l(x,y)\right]$ is simulated without applying FFT in order to avoid artefacts associated with FFT, and is calculated as $\mathrm{FT}\left[l(x,y)\right]=\sum_n \exp\left[-i\left(k_x x_n+k_y y_n\right)\right]$, where $(x_n, y_n)$ are the exact atomic coordinates, not pixels. $\mathrm{FT}^{-1}\left\{\mathrm{FT}\left[v_z(x,y)\right]\mathrm{FT}\left[l(x,y)\right]\right\}$ is calculated by applying an inverse FFT to the product of $\mathrm{FT}\left[v_z(x,y)\right]$ and $\mathrm{FT}\left[l(x,y)\right]$.

The exit wave after passing through the sample is given by the product of the incident wave $\psi_0(x,y)$ and the transmission function of the sample $u(x,y)=\psi_0(x,y)t(x,y)$. The CBED pattern is then simulated as the square of the amplitude of the FT of the exit wave. For a bilayer system, the transmission function of the bilayer was calculated as a product of transmission functions of individual layers.

## C. Sample preparation and acquisition

The samples were obtained by mechanical stacking of mechanically exfoliated graphene and hBN layers on a Si/$SiO_2$ substrate. By using "pick and lift" technique [12] - the whole stack was then transferred on a quantifoil carbon support film for observation in the TEM. CBED was realized in a probe side aberration corrected scanning TEM at an accellerating voltage of 80 keV and a convergence semiangle, $\alpha$, of ~6 – 8 mrad. During experiment, the focal lengths of the objective and condenser lenses were kept constant, thus there was no change in the convergence angle. The sample *z*-position was changed by moving the sample along the optical axis. Bilayer structure significantly improves the stability of 2D crystals upon exposure to high-energy electrons and the electron dose required for CBED imaging is low, so no evidence of knock-on damage was observed during prolonged data acquisition [13, 14].

## D. Analysis of experimental CBED patterns

For experimental images, it was observed that the noise level is decreasing towards higher scattering angles. To minimize the effect of the noise dependence on the scattering angle position, the signal and the noise were subtracted at a selected scattering angle, corresponding to the maximal extend of a CBED spot.

## Acknowledgements

The authors acknowledge support by the Ministry of Education (Singapore) through the Research Centre of Excellence program (Award EDUN C-33-18-279-V12, Institute for Functional Intelligent Materials), EU Graphene Flagship Program (contract CNECTICT-604391), European Research Council Synergy Grant Hetero2D (contract 319277), European Research Council Starting Grant EvoluTEM (contract 715502), the Royal Society (UK, grant number RSRP\R\190000), EPSRC grants EP/S019367/1, EP/P026850/1 and EP/N010345/1, EP/P009050/1, EP/S021531/1, FLAG-ERA project TRANS2DTMD, Swiss National Foundation Research Grant 200021_197107.